\documentclass[a4paper,12pt]{article}

\usepackage{jcappub} 
\usepackage[T1]{fontenc}
\usepackage[utf8]{inputenc}
\usepackage{xcolor}
\usepackage{soul}
\usepackage{cleveref}

\let\vec\mathbf
\newcommand{\es}[2] {\begin{equation} \label{#1} \begin{split} #2 \end{split} \end{equation}}
\newcommand{\mpl}{M_{\rm pl}}
\newcommand{\D}{{\rm d}}
\newcommand{\vk}{\vec{k}}
\newcommand{\fnl}{f_{\rm NL}}
\newcommand{\tfnl}{\tilde{f}_{\rm NL}}

\title{Probing Dark Matter Isocurvature with Primordial Non-Gaussianity}

\author[a]{Michael Geller,}
\author[b]{Soubhik Kumar,}
\author[c,d,e]{Lian-Tao Wang}
\affiliation[a]{School of Physics and Astronomy, Tel-Aviv University, Tel-Aviv 69978, Israel}
\affiliation[b]{Center for Cosmology and Particle Physics, Department of Physics,
New York University, New York, NY 10003, USA}
\affiliation[c]{Department of Physics, The University of Chicago, Chicago, IL 60637, USA}
\affiliation[d]{Enrico Fermi Institute, University of Chicago, Chicago, IL 60637, USA}
\affiliation[e]{Kavli Institute for Cosmological Physics, University of Chicago, Chicago, IL 60637, USA}
\emailAdd{micgeller@tauex.tau.ac.il}
\emailAdd{soubhik.kumar@nyu.edu}
\emailAdd{liantaow@uchicago.edu}

\begin{document}

\abstract{Multiple fields can become dynamical during the inflationary epoch. We consider an example where a light field acquires isocurvature fluctuations during inflation and contributes to the dark matter abundance at late times. Interactions between the light field and the adiabatic sector contribute to mixed adiabatic-isocurvature non-Gaussianity (NG). We show the resulting form of NG has a different kinematic dependence than the `local shape' commonly considered, and highlight the parameter space where a dedicated search is expected to significantly improve the current {\it Planck} sensitivity. We interpret our results in the context of the QCD axion and illustrate how the proposed NG searches can improve upon the existing searches for isocurvature power spectrum and bispectrum.}

\maketitle

\section{Introduction}

Non-Gaussianities (NG) in primordial density perturbations can reveal the nature of their dynamical origin and related physics (for reviews see~\cite{Bartolo:2004if, Chen:2010xka}). The search for, and the non-observation of NG in the correlations of both adiabatic and isocurvature perturbations~\cite{Planck:2019kim} have already helped inform us in early universe model building.
Among the various measures of NG, the three-point correlation function, i.e., the bispectrum, has been probed extensively, especially using the Cosmic Microwave Background (CMB) and also Large-Scale Structure (LSS) (e.g., \cite{eBOSS:2021jbt, Cabass:2022wjy, DAmico:2022gki}).
Assuming translational and rotational invariance, the primordial bispectrum depends only on the magnitudes $k_1, k_2, k_3$ of the three fluctuation modes which form a triangle, $\vec{k}_1 + \vec{k}_2+ \vec{k}_3=0$.
Different inflationary scenarios exhibit bispectra peaks for different `shapes' of the triangles.

Among these shapes, a primary focus has been on the bispectrum with the so-called `local shape'~\cite{Komatsu:2001rj}.
This is particularly relevant when multiple fields during inflation are present.
In such scenarios initial isocurvature fluctuations can convert into curvature fluctuations on superhorizon scales, generating the local shape NG in the process.
In particular, the curvaton scenario~\cite{Enqvist:2001zp,Lyth:2001nq,Moroi:2001ct} predict the local NG parameter $f_{\rm NL}^{\rm loc} \sim 1$~\cite{Lyth:2002my, Bartolo:2003jx, Sasaki:2006kq}, a target for various upcoming surveys.
Since the isocurvature perturbations are constrained to be subdominant compared to the adiabatic perturbations, the standard expectation is that the associated bispectrum, involving only adiabatic perturbations, would be the leading discovery channel for interesting dynamics. 

In this work, we point out that bispectrum involving {\it both} adiabatic and isocurvature perturbation can be indispensable in uncovering the fundamental dynamics leading to NG.
In particular, we consider an example in which the adiabatic density perturbation is generated by a curvaton \cite{Enqvist:2001zp,Lyth:2001nq,Moroi:2001ct}. At the same time, there is another light field, with an approximate shift symmetry, that contributes to the dark matter (DM) abundance. The interaction between the curvaton and the additional light field can come from a kinetic mixing, which preserves the shift symmetry and is straightforward to UV complete.  In this case, we show that correlations involving two adiabatic and one DM isocurvature mode provide the leading channel in observing the existence of the additional light field and proving such a coupling.\footnote{This model also predict sizable NG with more than one isocurvature perturbation. However, due to the limited sensitivities in the observations, these are less useful.}
Furthermore, the bispectrum involving only adiabatic fluctuations can also be larger than the minimal expectation $f_{\rm NL}^{\rm loc} \sim 1$. 

In addition, the predicted shape of NG in this class of models is not of the local type as considered in~\cite{Bartolo:2001cw, Kawasaki:2008sn, Langlois:2008vk, Hikage:2008sk, Kawasaki:2008pa, Langlois:2011hn, Hikage:2012tf, Langlois:2012tm, Montandon:2020kuk}, but rather of the type considered in, e.g.,~\cite{Creminelli:2003iq, Creminelli:2005hu, Senatore:2009gt} (in the context of adiabatic NG), mediated by derivative interactions.
We quantify the difference in shape following the approach of~\cite{Babich:2004gb}.
We find the `cosine'~\cite{Babich:2004gb} between the shapes employed in~\cite{Planck:2019kim} and the ones studied here is $\lesssim 0.5$, indicating a factor of few stronger sensitivity could be obtained by performing a dedicated analysis just by using the existing {\it Planck} data.
In addition, for certain values of local NG parameters, the `cosine'~\cite{Babich:2004gb} is $\sim 0.1$, implying an even better improvement would be feasible.
This highlights the importance of optimizing the template used in interpreting the relevant observations.

A natural application of the setup considered here is to a scenario in which the light field is the QCD axion.  We demonstrate that the bispectrum involving two adiabatic and one isocurvature mode is indeed the leading observable for the axion-curvaton interaction. In particular, it is more powerful than both the imprint of isocurvature power spectrum and the isocurvature-only bispectrum in revealing the existence of the axion sector.
We find a dedicated search would be able to probe novel regions of axion parameter space, unconstrained by the current stellar cooling and CMB bounds.

The rest of the paper is organized as follows. 
In Sec.~\ref{sec:limits} we summarize the shapes and the existing limits derived in the previous literature. 
This also sets up the notation we use in Sec.~\ref{sec:model} to describe the model.
We compute the predicted shapes and their overlap with the local shape to highlight the importance of a dedicated analysis.
We also interpret these results in the context of the QCD axion.
In Sec.~\ref{sec:general} we comment on other model-building possibilities. 
We also show how observable adiabatic-isocurvature NG can arise in the more model-agnostic effective field theory of inflation framework.
We conclude in Sec.~\ref{sec:con}.

\section{Notation and Existing Limits}\label{sec:limits}
In this section, we briefly summarize the existing searches for isocurvature bispectrum carried out by the Planck collaboration~\cite{Planck:2019kim}.
This will also establish the convention we use for the following discussion.
The templates used to search for isocurvature bispectrum is a generalized version of the template for local NG:
\es{eq:B_conv}{
B^{IJK}(k_1,k_2,k_3) & \equiv \langle \Phi^I(\vec{k}_1) \Phi^J(\vec{k}_2) \Phi^K(\vec{k}_3) \rangle'\\
& = 2 f_{\rm NL}^{I,JK} P_\Phi(k_2) P_\Phi(k_3) + 2 f_{\rm NL}^{J,KI} P_\Phi(k_1) P_\Phi(k_3) + 2 f_{\rm NL}^{K,IJ} P_\Phi(k_1) P_\Phi(k_2), 
}
where $\Phi(\vec{k})$ is the Bardeen potential~\cite{Bardeen:1980kt}.
The symbol $'$ denotes that momentum-conserving delta functions have been taken out, i.e., $\langle \cdots \rangle \equiv \langle \cdots \rangle'(2\pi)^3 \delta(\vec{k}_1+ \vec{k}_2 +\vec{k}_3)$.
The labels $I, J, K$ take values $a$ or $i$, and denote whether we have an adiabatic mode or an isocurvature mode, respectively.
The power spectrum is denoted as $P_\Phi(k) \equiv \langle \Phi(\vec{k})\Phi(-\vec{k})\rangle'$ with $k=|\vec{k}|$.
There is an exchange symmetry over the two indices after the comma in $f_{\rm NL}^{I, JK}$.
Therefore, for the parametrization in~\eqref{eq:B_conv}, six possible quantities determine the strength of NG: $f_{\rm NL}^{a, aa}$, $f_{\rm NL}^{a, ai}$, $f_{\rm NL}^{a, ii}$, $f_{\rm NL}^{i, aa}$, $f_{\rm NL}^{i, ai}$, $f_{\rm NL}^{i, ii}$.
The current observational constraint on these for DM isocurvature, from {\it Planck} 2018 temperature and polarization data, is given by~\cite{Planck:2019kim}
\es{}{
&\fnl^{a,aa} = -2.5 \pm 5.0,~~\fnl^{a,ai} = -10 \pm 10,~~\fnl^{a,ii} = -450\pm 520,\\ &\fnl^{i,aa} = 20 \pm 28,~~\fnl^{i,ai} = -32\pm 46,~~\fnl^{i,ii} = -290\pm 210.
}
Here each of the six parameters is analyzed independently, i.e., it is assumed that only one of these parameters is non-zero at a time.
However, for the scenario we describe later, all these parameters are expected to be non-zero.
A joint analysis, considering all the six parameters, is more appropriate and the constraints are given by~\cite{Planck:2019kim}
\es{eq:pl_joint}{
&\fnl^{a,aa} = 4 \pm 10,~~\fnl^{a,ai} = -14 \pm 21,~~\fnl^{a,ii} = -3100\pm 1500,\\ &\fnl^{i,aa} = 96 \pm 52,~~\fnl^{i,ai} = 190\pm 180,~~\fnl^{i,ii} = -640\pm 400.
}

To compare with theory predictions, it is convenient to relate the above parameters in terms NG defined in terms of primordial curvature $\zeta$ and isocurvature $S$ perturbations~\cite{Langlois:2011hn}:
\es{eq:B_tilde_conv}{
\tilde{B}^{ABC}(k_1,k_2,k_3) & \equiv \langle X^A(\vec{k}_1) X^B(\vec{k}_2) X^C(\vec{k}_3) \rangle'\\
& = \tilde{f}_{\rm NL}^{A,BC} P_\zeta(k_2) P_\zeta(k_3) +  \tilde{f}_{\rm NL}^{B,CA} P_\zeta(k_1) P_\zeta(k_3) +  \tilde{f}_{\rm NL}^{C,AB} P_\zeta(k_1) P_\zeta(k_2).
}
The indices $A, B, C$ can each be either $\zeta$ or $S$, and the fluctuation $X$ is accordingly determined, i.e., $X^\zeta \equiv \zeta$ and $X^S \equiv S$.
To relate $\fnl$ with $\tfnl$, we use the relations $\Phi = 3\zeta/5$ and $\Phi = S/5$, valid during matter domination, for superhorizon adiabatic and isocurvature modes, respectively.
Furthermore, in the {\it Planck} search, it is assumed that the power spectrum of $S$ has the same spectral index as that of $\zeta$.
Thus for the purpose of constraining isocurvature NG, any difference in the magnitude of $P_S$ and $P_\zeta$ is absorbed into the parameters $\tfnl$, and $P_S$ is taken to be equal to $P_\zeta$.
Using this, the numerical factors relating $\fnl$ with $\tfnl$ can be derived~\cite{Planck:2019kim}
\es{}{
\tfnl^{\zeta,\zeta\zeta} = (6/5)\fnl^{a,aa},~ \tfnl^{\zeta,\zeta S} = (2/5) \fnl^{a,ai},~\tfnl^{\zeta, S S} = (2/15) \fnl^{i,aa},\\
\tfnl^{S,\zeta \zeta} = (18/5)\fnl^{i,aa},~
\tfnl^{S, \zeta S} = (6/5) \fnl^{i,ai},~
\tfnl^{S, S S} = (2/5) \fnl^{i,ii}.
}
In terms of these, the {\it Planck} constraints in~\eqref{eq:pl_joint} read:
\es{eq:tfnl_bound}{
&\tfnl^{\zeta,\zeta\zeta} = 4.8\pm 12,~~\tfnl^{\zeta,\zeta S} = -5.6\pm 8.4,~~\tfnl^{\zeta,S S} = -413\pm 200,\\ &\tfnl^{S,\zeta \zeta} =  346 \pm 187,~~\tfnl^{S,\zeta S} = 228\pm 216,~~\tfnl^{S,SS} = -256 \pm 160.
}
Therefore, the current constraints on $\tfnl$ are ${\cal O}(100)$, except for $\tfnl^{\zeta,\zeta\zeta}$ and $\tfnl^{\zeta,\zeta S}$, for which the constraints are an order of magnitude stronger.
For the forecasted sensitivity of future surveys, such as LiteBIRD and CMB-S4, see Ref.~\cite{Montandon:2020kuk}.

While certain multifield models predict the bispectrum shape dictated by~\eqref{eq:B_tilde_conv}, it is not the only possibility.
We will show in the next section that in scenarios involving light, (pseudo)-Goldstone bosons, DM isocurvature NG can naturally arise with a shape different from the local shape~\eqref{eq:B_tilde_conv}. 
Such interactions are mediated by derivative couplings, and therefore the shape is peaked at the equilateral triangle configurations $k_1 \simeq k_2 \simeq k_3$.
Correspondingly, the overlap between the equilateral shape and the local shape in~\eqref{eq:B_tilde_conv} is small.
This implies that a search carried out using~\eqref{eq:B_tilde_conv} will not be optimal in detecting the actual equilateral shape.
To have a simple measure of this, the `cosine' between two shapes can be computed~\cite{Babich:2004gb}.
First, a dot product is defined:
\begin{align}
	B_1\cdot B_2 \equiv \sum_{\vec{k}_i} {\frac{\tilde{B}^{ABC}_1(k_1, k_2, k_3) \tilde{B}^{ABC}_2(k_1, k_2, k_3)}{\sigma_{k_1}^2 \sigma_{k_2}^2 \sigma_{k_3}^2}},
\end{align}
with $\sigma_{k}^2$ being the variance of a given $k$-mode and the sum running over all momentum triangle configurations.
The cosine is then defined as,
\es{eq:cosine}{
	\cos(B_1, B_2) \equiv \frac{B_1\cdot B_2}{(B_1\cdot B_1)^{1/2} (B_2\cdot B_2)^{1/2}}.
}
A cosine value close to unity implies a template designed to detect shape $B_1$ would also be optimal in detecting NG which has shape $B_2$, while a cosine value closer to zero would imply the opposite.
We will compute the associated cosines in Sec.~\ref{sec:model}.

\section{Model of Dark Matter Isocurvature}
\label{sec:model}
The cosmological evolution we consider is similar to the curvaton scenario~\cite{Enqvist:2001zp,Lyth:2001nq,Moroi:2001ct}.
While this is not the most minimal scenario, we argue in Sec.~\ref{sec:general} why the standard slow-roll inflationary scenario, in the absence of any curvaton dynamics, gives a suppressed adiabatic-isocurvature NG.
On the other hand, in Sec.~\ref{sec:general} we also show that the EFT of inflation framework~\cite{Cheung:2007st} can give observable NG, while being agnostic about the full homogeneous dynamics.
From that perspective, the model in this section can be considered a `UV completion' of the EFT description and the EFT cutoff scale can be explained in terms of the fundamental UV parameters.
\subsection{Setup}
Consider an inflaton field $\phi$ that gives rise to the homogeneous spacetime expansion during inflation. 
The fluctuations of $\phi$ are much smaller than $\sim$$10^{-5}$, the typical size of CMB anisotropies.
Instead, these anisotropies originate from a `curvaton' field $\sigma$ which behaves as a spectator field during inflation, having a subdominant energy density compared to $\phi$.
In the post-inflationary universe, $\sigma$ eventually dominates the total energy density, and fluctuations of $\sigma$ determine the primordial density fluctuations.
Eventually, $\sigma$ decays and gives rise to the standard radiation-dominated universe with the fluctuations of $\sigma$ determining the CMB anisotropies.
The dimensionless primordial scalar power spectrum, originating from $\sigma$, is given by
\es{}{
{\cal P}_\zeta(k) \equiv {k^3 \over 2\pi^2} \langle \zeta(\vec{k})\zeta(\vec{-k})\rangle' \equiv {k^3 \over 2\pi^2} P_\zeta(k) = {4 \over 9}{H^2 \over 4\pi^2 \sigma_0^2},
} 
where $H$ is the Hubble scale and $\sigma_0$ is the `misaligned' field value of $\sigma$ when the $k$-mode exits the horizon during inflation.
The tilt is given by,
\es{eq:tilt}{
n_s-1 =-2\epsilon +{2\over 3}\eta_\sigma,
}
where $\epsilon = -\dot{H}/H^2$ and $\eta_\sigma = m_\sigma^2/H^2$ is given by the mass of the curvaton $m_\sigma$.
The tensor-to-scalar ratio ${\rm r}$ is determined by
\es{}{
{\rm r} = {8 H^2/\mpl^2 \over 4H^2/(9\sigma_0^2)} = 18 \sigma_0^2 / \mpl^2.
}
In particular, different from the single field scenario, ${\rm r}$ is not determined by $\epsilon$, since the scalar fluctuations originate from a field that is different from the one that drives the homogeneous expansion.

We now consider the effect of another scalar field $\chi$, that gives (a subcomponent of) the DM abundance, on this cosmology.
In this subsection, we remain agnostic about the particle nature of $\chi$ and only impose that it is a light field during inflation, obeying an approximate shift symmetry: $\chi \rightarrow \chi + {\rm constant}$.
Later in this section, we consider the case where $\chi$ is the QCD axion~\cite{Peccei:1977hh,Peccei:1977ur,Weinberg:1977ma,Wilczek:1977pj} that solves the strong {\it CP} problem and also contributes to the DM abundance~\cite{Preskill:1982cy,Abbott:1982af,Dine:1982ah}.
In both these scenarios, we treat $\chi$ as effectively massless during inflation and it only starts to oscillate when the Hubble scale falls below the mass of $\chi$, after which $\chi$ abundance dilutes as matter.
In our scenario, this happens after $\sigma$ decay such that $\chi$ never dominates the energy density of the universe before the matter-radiation inequality.
The total curvature perturbation $\zeta$ after inflation on superhorizon scales is given by,
\es{}{
\zeta = f_r \zeta_r + f_\sigma \zeta_\sigma + f_\chi \zeta_\chi,
}
such that $\sum_{i=r,\sigma,\chi} f_i = 1$ with $f_i = \dot{\rho}_i/\dot{\rho}_{\rm tot}$ and $\rho_{\rm tot} = \sum_{i=r,\sigma,\chi} \rho_i$ is the total energy density.
$\zeta_i$ denotes the individual gauge-invariant perturbation in each species, for a review see~\cite{Malik:2008im}.
The label $r$ denotes radiation decay products of the inflaton.
As in the curvaton paradigm, we assume $\zeta_r\ll \zeta_\sigma \sim 10^{-5}$.
Furthermore, the fractional energy density in inflaton decay products, $f_r$, decreases with time since both $\sigma$ and $\chi$ dilute slower than radiation.
Therefore, after $\sigma$ comes to dominate the energy density, we can approximate
\es{}{
\zeta \approx (1-f_\chi) \zeta_\sigma + f_\chi \zeta_\chi = \zeta_\sigma + {1\over 3}f_\chi S_\chi,
}
where $f_\chi \approx 1-f_\sigma$ and $S_\chi = 3(\zeta_\chi - \zeta_\sigma)$ denotes the isocurvature fluctuation of $\chi$ relative to $\sigma$. 
$S_\chi$ is given by $ (\delta\rho_\chi/\rho_\chi - \delta\rho_\sigma/\rho_\sigma) \approx \delta\rho_\chi/\rho_\chi$ where we have used the isocurvature initial condition $\delta\rho_\chi + \delta\rho_\sigma = 0$ along with the fact that $\rho_\sigma \gg \rho_\chi$ much before matter-radiation equality.
We can write $S_\chi \approx 2\delta\chi /\chi_0$, where $\chi_0$ is the homogeneous field value of the $\chi$ field.
The power spectrum of $S_\chi$ is then given by
\es{eq:Ps_chi}{
{\cal P}_{S_\chi} = 
\left({H \over \pi \chi_0}\right)^2.
}
If $\chi$ is the QCD axion then $\chi_0 = \theta_i F_a$ where $\theta_i$ is the initial misalignment angle and $F_a$ is the axion decay constant.

\subsection{Interactions}
To describe the interaction between $\chi$ and $\sigma$, we impose a (softly-broken) shift symmetry on the system.
In the UV, such a shift symmetry can originate from gauge invariance, as we argue later.
We also demand a consistent EFT expansion such that higher order terms that we do not consider contribute subdominantly, and the derivative expansion is under control.
The leading interactions are then given by
\es{eq:leading_lag}{
{\cal L} = -{1\over 2}(\partial \sigma)^2 - {1\over 2}(\partial \chi)^2 - {1\over 2}(\partial s)^2 +\varepsilon \partial\sigma \partial \chi-{1\over 2} m^2_s s^2-(\partial \chi)^2 {s \over v_s} +\cdots.
}
Here and below, we use the shorthand notation $(\partial\sigma)^2 \equiv \partial_\mu\sigma \partial^\mu\sigma$, $\partial\sigma\partial\chi \equiv \partial_\mu\sigma \partial^\mu\chi$, and similarly for other quantities.
We have included a mass term for $\sigma$, as needed to explain the tilt of the scalar power spectrum, while $\chi$ is massless.
We also have a kinetic mixing $\varepsilon$ between $\chi$ and $\sigma$, and have kept the radial mode $s$ associated with $\chi$, defined as $\Sigma = (s+v_s)\exp(i\chi/v_s)/\sqrt{2}$.
One can remove the kinetic mixing by performing a field redefinition,
\es{eq:shift}{
\chi \rightarrow \chi + \varepsilon \sigma.
}
This field shift gives rise to higher dimensional derivative mixing terms,
\es{}{
(\partial \chi)^2 {s \over v_s} \rightarrow (\partial \chi)^2 {s \over v_s} + 2\varepsilon (\partial \chi) (\partial \sigma) {s \over v_s} + \varepsilon^2 (\partial \sigma)^2 {s \over v_s}.
}
The last term above gives a subdominant correction to the curvaton kinetic term when $s$ is set to some homogeneous value $\lesssim v_s$.
Assuming the radial mode mass $m_s > H$, we can integrate it out,\footnote{See~\cite{Chen:2023txq} for a complementary study where the effects of on-shell dynamics of $s$ were studied. Such effects give oscillatory adiabatic-isocurvature NG as in the context of cosmological collider physics~\cite{Chen:2009zp, Arkani-Hamed:2015bza}.} and that gives rise to various dimension-8 local interactions such as,
\es{eq:dim8_full}{
{1\over 2 m_s^2 v_s^2}\left(
(\partial \chi)^4 + 4 \varepsilon (\partial \chi)^2 (\partial \chi \partial \sigma)  + 2 \varepsilon^2 (\partial \chi)^2 (\partial \sigma)^2 \right.\\ \left.+ 4\varepsilon^2(\partial \chi\partial\sigma)^2 + 4  \varepsilon^3 (\partial \sigma)^2 (\partial \chi\partial \sigma)+ \varepsilon^4 (\partial \sigma)^4 \right).
}
The homogeneous part of the curvaton $\sigma_0$, which follows the EOM $\dot{\sigma}_0 \approx -m_\sigma^2 \sigma_0/(3 H)$, gives correction to the leading order Lagrangian considered in~\eqref{eq:leading_lag}.\footnote{Owing to the negligible mass of $\chi$, similar corrections coming from $\dot{\chi}_0$ can be neglected.}
For consistency, we demand such corrections to be parametrically small.
\begin{itemize}
	\item Correction to curvaton kinetic term:
	\es{}{
{\varepsilon^4 \dot{\sigma}_0^2 \over m_s^2 v_s^2} < 1.
}
\item Contribution to the speed of curvaton fluctuation:
\es{}{
c_\sigma^2 = \left(1 + {4\dot{\sigma}_0^2\varepsilon^4 \over m_s^2 v_s^2}\right)^{-1} \approx 1.
}
\item Correction to $\chi$ kinetic term:
\es{eq:eps2}{
{\varepsilon^2\dot{\sigma}_0^2 \over m_s^2 v_s^2} < 1.
}
\item Contribution to the speed of $\chi$ fluctuation:
\es{eq:speed_chi}{
c_\chi^2 = \left(1 + {4\dot{\sigma}_0^2\varepsilon^2 \over m_s^2 v_s^2}\right)^{-1} \approx 1.
}
\item Correction to curvaton-$\chi$ mixing:
\es{eq:kin_mix}{
{2\varepsilon^3\dot{\sigma}_0^2 \over m_s^2 v_s^2} < \varepsilon.
}
\end{itemize}
Therefore, the most stringent among these requirements is given by~\eqref{eq:eps2},~\eqref{eq:speed_chi}, and~\eqref{eq:kin_mix} with $\varepsilon < 1$. We can summarize these restriction by defining $\kappa \equiv \varepsilon^2 \dot{\sigma}_0^2/(m_s^2 v_s^2)$ and imposing $\kappa < 1$.

\paragraph{Origin of kinetic mixing.}
Eq.~\eqref{eq:leading_lag} describes kinetic mixing $\varepsilon$ between two light fields $\sigma$ and $\chi$.
From a UV perspective, $\varepsilon$ can arise in several ways, including a scenario involving one flat extra spatial dimension.
Consider two $U(1)$ gauge fields, $A_M$, and $B_M$, with gauge couplings $g_{5,A}$ and $g_{5,B}$, and field strengths $F_{MN}$ and $G_{MN}$, respectively.
The 5D Lagrangian is given by
\es{}{
{\cal L}_{5 \rm D} = -{1\over 4g_{5,A}^2}\int \D^5 x \sqrt{-g_{5D}} F^{MN}F_{MN} - {1\over 4g_{5,B}^2}\int \D^5 x \sqrt{-g_{5D}} G^{MN}G_{MN} \\- {\varepsilon_5 \over g_{5,A} g_{5,B}} \int \D^5 x \sqrt{-g_{5D}} F^{MN}G_{MN}+\cdots.
}
The gauge kinetic mixing $\varepsilon_5$ can arise by integrating out massive 5D fermions that are charged under both the $U(1)$'s.
To reduce to a 4D theory, we can choose orbifold boundary conditions such that only $A_5$ and $B_5$ have zero modes, while $A_\mu$ and $B_\mu$ do not.
Then the low-energy 4D Lagrangian is given by
\es{}{
{\cal L}_{4 \rm D,~eff} = -{2\pi R\over 2g_{5,A}^2}\int \D^4 x \sqrt{-g_{4D}} (\partial_\mu A_5)^2 -{2\pi R\over 2g_{5,B}^2}\int \D^4 x \sqrt{-g_{4D}} (\partial_\mu B_5)^2 \\- {4\pi R \varepsilon_5 \over g_{5,A} g_{5,B}} \int \D^4 x \sqrt{-g_{4D}} (\partial^\mu A_5)(\partial_\mu B_5)+\cdots.
}
Canonically normalizing the $A_5$ and $B_5$ kinetic terms, we can rewrite the above as the first three terms of~\eqref{eq:leading_lag} by identifying $A_5$ and $B_5$ with $\sigma$ and $\chi$, respectively, along with $\varepsilon = 2\varepsilon_5$. The shift symmetries in the 4D effective field theory come from the 5D gauge transformations of $A_5$ and $B_5$.

\subsection{Magnitude of Bispectrum}
\begin{figure}
	\begin{center}
		\includegraphics[width=0.6\textwidth]{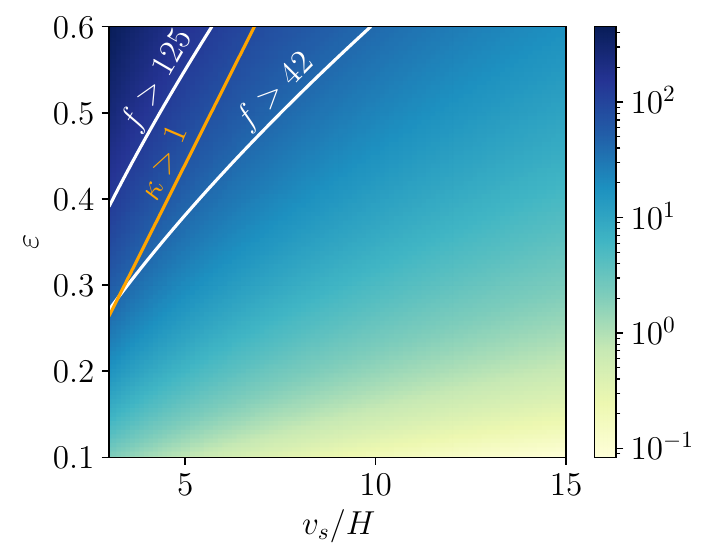}
	\end{center}
	\caption{Absolute size of $\tilde{f}_{\rm NL, equi}^{\zeta\zeta S}$. We fix $m_s = 2H$, $m_\sigma^2 = 0.03 H^2$, and saturate the current isocurvature power spectrum bound $\beta_{\rm iso}=0.038$. In the region above the orange line $\kappa > 1$ and the corrections discussed in~\eqref{eq:eps2}--\eqref{eq:kin_mix} become important. We can roughly recast the current constraint on this parameter space and that rules out the region above the upper white line, labeled `$f>125$', where $\tilde{f}_{\rm NL, equi}^{\zeta\zeta S} > 125$. A dedicated template would improve this bound by roughly a factor of 3, and thereby be sensitive to the region above the lower white line labeled `$f>42$' (see Sec.~\ref{sec:shape_bispec} for more details).}
 \label{fig_size}
\end{figure}

Eq.~\eqref{eq:dim8_full} determines the cubic interaction vertices between $\sigma$ and $\chi$.
This can be obtained by expressing the fields in terms of a homogeneous background and a fluctuating piece: $\sigma(t,\vec{x}) = \sigma_0(t) + \delta\sigma(t,\vec{x})$ and $\chi(t,\vec{x}) = \chi_0(t) + \delta\chi(t,\vec{x})$.
Practically, $\dot{\chi}_0 \approx 0$ since $\chi$ is approximately massless during inflation, and therefore we will use $\partial_\mu\delta\chi(t,\vec{x}) \approx \partial_\mu\chi(t,\vec{x})$. 
With that the cubic interactions are given by,
\es{eq:lag_cubic}{
{\cal L} = -{\dot{\sigma}_0 \over m_s^2 v_s^2} \left( 2\varepsilon (\partial\chi)^2\dot{\chi} + 2\varepsilon^2 \dot{\delta\sigma}(\partial\chi)^2 + 4 \varepsilon^2 (\partial\chi \partial\delta\sigma)\dot{\chi}\right.\\ \left.+ 2\varepsilon^3(\partial\delta\sigma)^2\dot{\chi} + 4\varepsilon^3\dot{\delta\sigma} (\partial\chi \partial\delta\sigma) + 2\varepsilon^4\dot{\delta\sigma}(\partial\delta\sigma)^2\right)+\cdots.
}
These determine the various types of mixed adiabatic-isocurvature bispectra.
The associated shapes have been computed in~\cite{Freytsis:2022aho} which we can express in terms of the parameters from~\eqref{eq:lag_cubic}.
To that end, we define a kinematic function
\es{}{
Q(\vec{k}_1, \vec{k}_2, \vec{k}_3) = {1 \over k_1^3 k_2^3 k_3^3}\left(\frac{2k_1^2 k_2^2 k_3^2}{k_{123}^3} - \frac{(\vk_2\cdot \vk_3) k_1^2}{k_{123}} \left(1 + \frac{k_{23}}{k_{123}} + \frac{2k_2 k_3}{k_{123}^2}\right)\right),
}
where $k_i = |\vec{k}_i|$ and $k_{ij\cdots m} = |\vec{k}_i|+|\vec{k}_j|+\cdots+|\vec{k}_m|$.
We also define $Q_{\rm full}(\vec{k}_1, \vec{k}_2, \vec{k}_3) = Q(\vec{k}_1, \vec{k}_2, \vec{k}_3) + Q(\vec{k}_2, \vec{k}_3, \vec{k}_1) + Q(\vec{k}_3, \vec{k}_1, \vec{k}_2)$.
The various three-point functions can then be expressed as
\es{eq:3pt_shapes}{
\langle \sigma(\vec{k}_1) \sigma(\vec{k}_2) \sigma(\vec{k}_3) \rangle' &= {\varepsilon^4 \dot{\sigma}_0 \over m_s^2 v_s^2} Q_{\rm full}(\vec{k}_1, \vec{k}_2, \vec{k}_3),\\
\langle \chi(\vec{k}_1) \sigma(\vec{k}_2) \sigma(\vec{k}_3) \rangle' &= {\varepsilon^3 \dot{\sigma}_0 \over m_s^2 v_s^2} Q_{\rm full}(\vec{k}_1, \vec{k}_2, \vec{k}_3),\\
\langle \chi(\vec{k}_1) \chi(\vec{k}_2) \sigma(\vec{k}_3) \rangle' &= {\varepsilon^2 \dot{\sigma}_0 \over m_s^2 v_s^2} Q_{\rm full}(\vec{k}_1, \vec{k}_2, \vec{k}_3),\\
\langle \chi(\vec{k}_1) \chi(\vec{k}_2) \chi(\vec{k}_3) \rangle' &= {\varepsilon \dot{\sigma}_0 \over m_s^2 v_s^2} Q_{\rm full}(\vec{k}_1, \vec{k}_2, \vec{k}_3),
}
where $\sigma(\vec{k})$ and $\chi(\vec{k})$ are Fourier transforms of $\delta\sigma (t,\vec{x}) $ and $\delta\chi (t,\vec{x}) $, respectively.
To convert the above into correlation function between $\zeta$ and $S$, we can use $\zeta = (2/3) (\delta \sigma/\sigma_0)$ and $S = f_{\rm DM} S_\chi = 2 f_{\rm DM} (\delta\chi/\chi_0)$, where $f_{\rm DM}$ is the fraction of DM abundance determined by $\chi$.\footnote{The DM isocurvature is defined as $S = 3(\zeta_{\rm DM} -\zeta_\gamma)$. Writing $\zeta_{\rm DM} = f_{\rm DM} \zeta_\chi + (1-f_{\rm DM})\zeta_c$, where $\zeta_c$ is the fluctuation in the other (adiabatic) component of DM, and setting $\zeta_c=\zeta_\gamma$, we get $S = f_{\rm DM}S_\chi$.}
While the shapes in~\eqref{eq:3pt_shapes} are different from the ones used in the {\it Planck} search~\eqref{eq:B_tilde_conv}, we can make a rough comparison by defining the following quantities~\cite{Freytsis:2022aho}:
\es{eq:new_fnl}{
\tilde{f}^{\zeta\zeta\zeta}_{\rm NL,equi} \equiv & {\langle \zeta(\vec{k}) \zeta(\vec{k}) \zeta(\vec{k}) \rangle' \over P_\zeta(k)^2} = {7 \over 6}{81 \sigma_0^4 \over 4H^4}\left({2H \over 3\sigma_0}\right)^3{\varepsilon^4 H^2 \dot{\sigma}_0 \over m_s^2 v_s^2},\\
\tilde{f}^{\zeta\zeta S}_{\rm NL,equi} \equiv &{\langle \zeta(\vec{k}) \zeta(\vec{k}) S(\vec{k}) \rangle' \over P_\zeta(k)^2} = {7 \over 6}{81 \sigma_0^4 \over 4H^4}\left({2H \over 3\sigma_0}\right)^2\left(2H f_{\rm DM} \over \chi_0\right){\varepsilon^3 H^2\dot{\sigma}_0 \over m_s^2 v_s^2},\\
\tilde{f}^{\zeta SS}_{\rm NL,equi} \equiv &{\langle \zeta(\vec{k}) S(\vec{k}) S(\vec{k}) \rangle' \over P_\zeta(k)^2} = {7 \over 6}{81 \sigma_0^4 \over 4H^4}\left({2H \over 3\sigma_0}\right)\left(2 H f_{\rm DM} \over \chi_0\right)^2{\varepsilon^2 H^2\dot{\sigma}_0 \over m_s^2 v_s^2},\\
\tilde{f}^{SSS}_{\rm NL,equi} \equiv &{\langle S(\vec{k}) S(\vec{k}) S(\vec{k}) \rangle' \over P_\zeta(k)^2} = {7 \over 6}{81 \sigma_0^4 \over 4H^4}\left(2H f_{\rm DM} \over \chi_0\right)^3{\varepsilon H^2\dot{\sigma}_0 \over m_s^2 v_s^2}.
}
To compare with observational bounds on isocurvature power spectrum, we define
\es{}{
\beta_{\rm iso} \equiv {P_S \over P_\zeta + P_S} \Rightarrow {\beta_{\rm iso} \over 1- \beta_{\rm iso}} = {P_S \over P_\zeta} = {9 \sigma_0^2 f_{\rm DM}^2 \over \chi_0^2}.
}
In terms of $\beta_{\rm iso}$,~\eqref{eq:new_fnl} reads:
\es{}{
\tilde{f}^{\zeta\zeta\zeta}_{\rm NL,equi} &= {7 H\dot{\sigma}_0 \varepsilon^4 \sigma_0 \over m_s^2 v_s^2} = -{21 H^2 \kappa \varepsilon^2 \over m_\sigma^2},\\
\tilde{f}^{\zeta\zeta S}_{\rm NL,equi} &= {7 H\dot{\sigma}_0 \varepsilon^3 \sigma_0 \over m_s^2 v_s^2}\left( {\beta_{\rm iso} \over 1- \beta_{\rm iso}} \right)^{1/2}= -{21 H^2 \kappa \varepsilon \over m_\sigma^2}\left( {\beta_{\rm iso} \over 1- \beta_{\rm iso}} \right)^{1/2},\\
\tilde{f}^{\zeta SS}_{\rm NL,equi} &= {7 H\dot{\sigma}_0 \varepsilon^2 \sigma_0 \over m_s^2 v_s^2} \left( {\beta_{\rm iso} \over 1- \beta_{\rm iso}} \right) = -{21 H^2 \kappa \over m_\sigma^2} \left( {\beta_{\rm iso} \over 1- \beta_{\rm iso}} \right),\\
\tilde{f}^{SSS}_{\rm NL,equi} &= {7 H\dot{\sigma}_0 \varepsilon \sigma_0 \over m_s^2 v_s^2}\left( {\beta_{\rm iso} \over 1- \beta_{\rm iso}} \right)^{3/2} = -{21 H^2\kappa \over \varepsilon m_\sigma^2}\left( {\beta_{\rm iso} \over 1- \beta_{\rm iso}} \right)^{3/2},
}
where $\kappa = \varepsilon^2 \dot{\sigma}_0^2/(m_s^2 v_s^2) < 1$ to obey the constraints~\eqref{eq:eps2}--\eqref{eq:kin_mix}.
We have used the homogeneous EOM for $\sigma$, $3 H \dot{\sigma}_0 \approx -m_\sigma^2 \sigma_0$.
To obtain an estimate of the magnitude of NG, we choose the following benchmark parameter values: $m_\sigma^2 = 0.03H^2$, $\varepsilon =0.3$, $v_s = 10H$, and $m_s = 2H$.
We also use $H/\sigma_0 = 4.3\times 10^{-4}$, determined by the scalar power spectrum, and $\beta_{\rm iso} = 0.038$, current upper limit on the magnitude of the DM isocurvature power spectrum~\cite{Planck:2018jri}.
These parameter choices imply $\kappa \approx 0.1$ and 
\es{eq:tfnl_equi}{
\tilde{f}^{\zeta\zeta\zeta}_{\rm NL,equi} \approx -7.7, \tilde{f}^{\zeta\zeta S}_{\rm NL,equi} \approx -5.1, \tilde{f}^{\zeta SS}_{\rm NL,equi} \approx -3.4, \tilde{f}^{SSS}_{\rm NL,equi} \approx -2.2.
}
Although the bounds in~\eqref{eq:tfnl_bound} were derived using a different shape, we can make a rough comparison using~\eqref{eq:tfnl_equi}. 
For example, the Planck constraint~\eqref{eq:tfnl_bound} gives $\tfnl^{\zeta,\zeta S} = -5.6\pm 8.4$, comparable to the values obtained in~\eqref{eq:tfnl_equi}.
We note that the NG involving two adiabatic and one isocurvature perturbation, $\tilde{f}^{\zeta\zeta S}_{\rm NL,equi}$, is the leading probe of DM isocurvature NG.
While $\tilde{f}^{\zeta SS}_{\rm NL,equi}$ and $\tilde{f}^{SSS}_{\rm NL,equi}$ are not too different in strengths from $\tilde{f}^{\zeta\zeta S}_{\rm NL,equi}$, the observational bounds on NG involving more than one isocurvature perturbation is an order of magnitude weaker~\eqref{eq:tfnl_bound}.
Therefore, in the following, we focus on the correlation function $\langle \zeta \zeta S\rangle$ to analyze the difference in shapes between~\eqref{eq:3pt_shapes} and the ones considered in the {\it Planck} search~\eqref{eq:B_tilde_conv}.
Similar steps can be repeated for models where $\langle \zeta S S\rangle$ and $\langle SS S\rangle$ are relevant.
In Fig.~\ref{fig_size} we show the absolute size of $\tilde{f}_{\rm NL, equi}^{\zeta\zeta S}$ as a function of $v_s/H$ and $\varepsilon$.

\subsection{Shape of Bispectrum}
\label{sec:shape_bispec}
\begin{figure}
	\begin{center}
\includegraphics[width=0.45\textwidth]{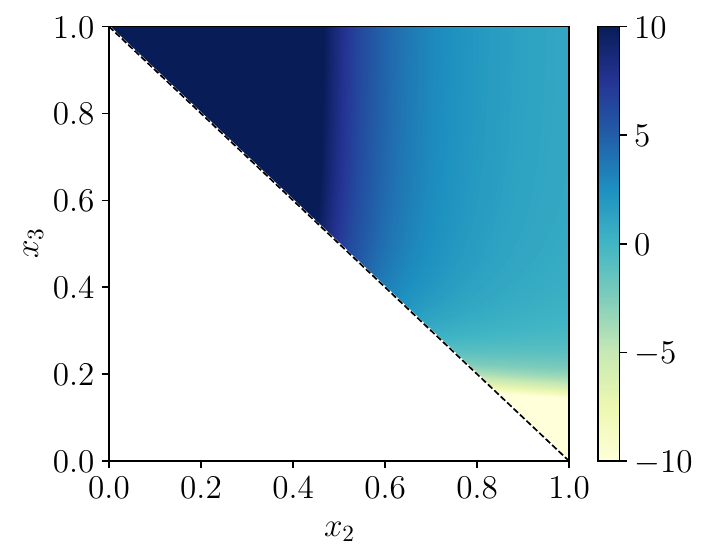}
\includegraphics[width=0.45\textwidth]{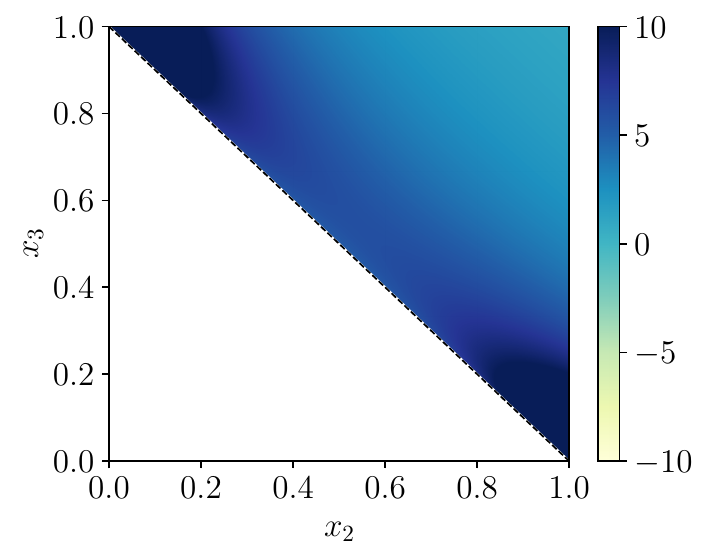}
	\end{center}
	\caption{{\it Left.} Local shape~\eqref{eq:shapes} for $\tfnl^{\zeta,\zeta S} = -5.6$ and $\tfnl^{S,\zeta\zeta}=346$, as per the central values in~\eqref{eq:tfnl_bound}. {\it Right.} Derivative shape~\eqref{eq:shapes}. We have imposed the triangle inequality $x_2+x_3\geq 1$. Here $x_2 = k_2/k_1$ and $x_3 = k_3/k_1$ and both the bispectra are normalized to unity at $x_2 = x_3=1$. The two shapes are significantly different from each other, as further quantified via the cosine in Fig.~\ref{fig_cosine}.}
 \label{fig_shape}
\end{figure}

In Fig.~\ref{fig_shape}, we compare the two classes of shapes given by~\eqref{eq:B_tilde_conv}  and~\eqref{eq:3pt_shapes}.
For convenience, we rewrite the shapes by ignoring the overall prefactors which are not relevant for examining the shapes,
\es{eq:shapes}{
\langle \zeta(\vec{k}_1)\zeta(\vec{k}_2) S(\vec{k}_3)\rangle'_{\rm local} &\propto \tfnl^{\zeta,\zeta S} \left({1\over k_2^3}{1\over k_3^3} + {1\over k_1^3}{1\over k_3^3}\right) + \tfnl^{S,\zeta \zeta}{1\over k_1^3}{1\over k_2^3},\\
\langle\zeta(\vec{k}_1)\zeta(\vec{k}_2) S(\vec{k}_3)\rangle'_{\rm derivative} &\propto Q(\vec{k}_1, \vec{k}_2, \vec{k}_3) + Q(\vec{k}_2, \vec{k}_3, \vec{k}_1) + Q(\vec{k}_3, \vec{k}_1, \vec{k}_2).
}
The local shape is symmetric under $\vec{k_1}\leftrightarrow \vec{k_2}$ and is only dependent on the ratio $\tfnl^{\zeta,\zeta S}/\tfnl^{S,\zeta \zeta}$.
On the other hand, the derivative shape is symmetric under the cyclic permutation of $\vec{k_1}, \vec{k_2}, \vec{k_3}$.
To compare the shapes we choose $k_1$ to be the largest side of the triangle and define the ratios $x_2 = k_2/k_1$ and $x_3=k_3/k_1$.
Since at the moment we are interested in the shape, not the overall magnitude, we normalize both the shapes to be unity at $x_2=x_3=1$.
We also restrict the shapes in the region $x_2+x_3\geq 1$ (triangle inequality).
From Fig.~\ref{fig_shape} we note that the local shape is not symmetric under the exchange $x_2\leftrightarrow x_3$, while the derivative shape is, as expected from~\eqref{eq:shapes}. 
For the chosen parameter points, the two shapes are significantly different.

\begin{figure}
	\begin{center}
\includegraphics[width=0.48\textwidth]{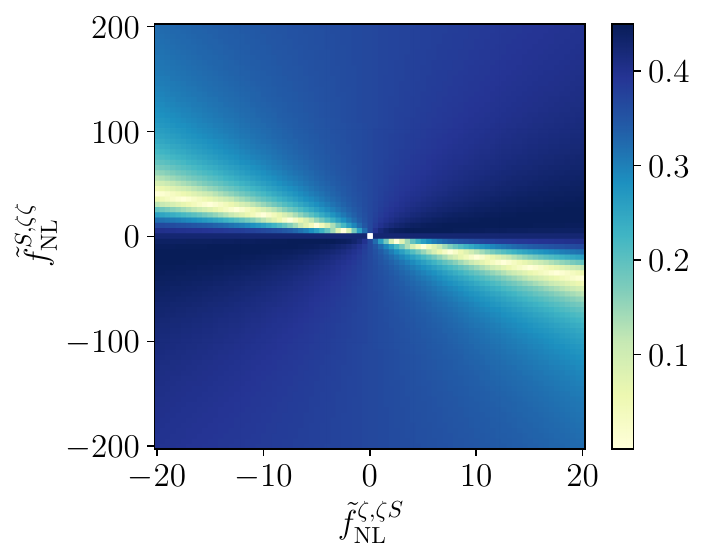}
\includegraphics[width=0.48\textwidth]{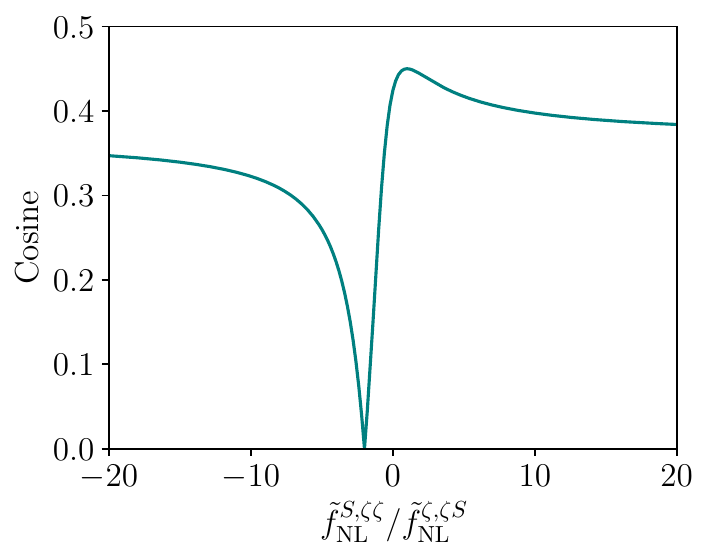}
	\end{center}
	\caption{{\it Left.} Cosine between the local shape and the derivative shape~\eqref{eq:shapes} for a range of $\tfnl^{\zeta,\zeta S}$ and $\tfnl^{S,\zeta \zeta}$ motivated by~\eqref{eq:tfnl_bound}. Changing the signs of both $\tfnl^{\zeta,\zeta S}$ and $\tfnl^{S,\zeta \zeta}$ gives the same absolute value of the cosine. Rescaling both $\tfnl^{\zeta,\zeta S}$ and $\tfnl^{S,\zeta \zeta}$ by the same factor also gives the same value of the cosine, also seen in the above plot. {\it Right.} These features are utilized in the right panel where we show the cosine as a function of the ratio $\tfnl^{S,\zeta \zeta}/\tfnl^{\zeta,\zeta S}$. We see that by using the correct shape, we can improve the sensitivity by at least a factor of two in comparison with a search treating  $\tfnl^{\zeta,\zeta S}$ and $\tfnl^{S,\zeta \zeta}$ as free parameters. At the same time, there could be one order of magnitude difference in sensitivity for certain choices of $\tfnl^{\zeta,\zeta S}$ and $\tfnl^{S,\zeta \zeta}$. 
}
 \label{fig_cosine}
\end{figure}

To quantify the difference between the two shapes, we can compute the cosine between them as per~\eqref{eq:cosine} for a range of values of $\tfnl^{\zeta,\zeta S}$ and $\tfnl^{S,\zeta \zeta}$.
The result is shown in Fig.~\ref{fig_cosine} for the absolute value of the cosine.
The cosine depends only on the ratio $\tfnl^{S,\zeta \zeta}/\tfnl^{\zeta,\zeta S}$, as evident from~\eqref{eq:shapes}.
In particular, the result is symmetric for $(\tfnl^{\zeta,\zeta S}, \tfnl^{S,\zeta \zeta}) \rightarrow -(\tfnl^{\zeta,\zeta S}, \tfnl^{S,\zeta \zeta})$.
We utilize this feature to show the cosine as a function of this ratio in the right panel of Fig.~\ref{fig_cosine}.
The minimum value of the cosine is reached for $\tfnl^{S,\zeta \zeta}=-2\tfnl^{\zeta,\zeta S}$.
For this value the local shape~\eqref{eq:shapes} vanishes in the equilateral limit.
We see for the entire range of the parameter space, the cosine is smaller than $0.5$ in magnitude.
In fact, for regions of parameter space, the cosine can be $\sim0.1$.
This implies, even without including new data, performing a reanalysis with the dedicated template can improve the sensitivity by an order of magnitude (inverse of the cosine). 

We can also do a very rough recast of the existing bound by calculating the (3D) `fudge factor' as in~\cite{Babich:2004gb} which accounts for both the cosine and the norm of the signal.\footnote{Note, a better recast can be obtained by computing the 2D fudge factors~\cite{Babich:2004gb}, as well as, including the correlation between $\tfnl^{\zeta,\zeta S}$ and $\tfnl^{S,\zeta \zeta}$.}
For $\tfnl^{\zeta,\zeta S}=-5.6$ and $\tfnl^{S,\zeta \zeta}=346$, the fudge factor is $\approx 0.067$ and the cosine is $\approx 0.36$.
This means the current constraint on $\tilde{f}_{\rm NL, equi}^{\zeta\zeta S}$, based on~\eqref{eq:tfnl_bound}, is roughly $\sim 8.4/0.067 \approx 125$.
Including the derivative shape template would tighten this bound by a factor of the inverse cosine, $\sim 3$.
We illustrate this improvement in Fig.~\ref{fig_size}.

\subsection{Interpretation for the QCD Axion}

\begin{figure}[h!]
	\begin{center}
\includegraphics[width=0.7\textwidth]{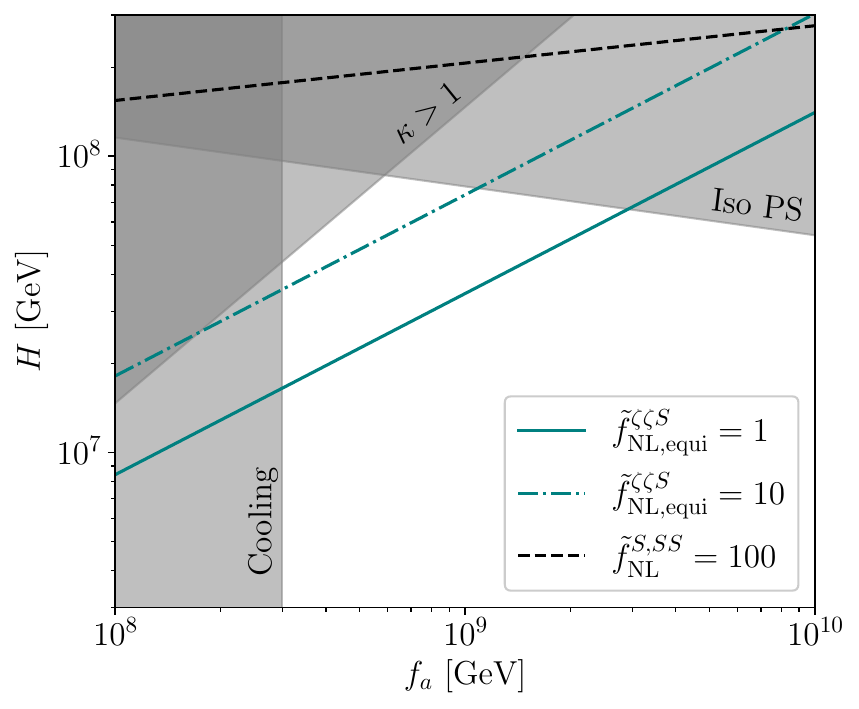}
	\end{center}
	\caption{QCD Axion parameter space as a function of the decay constant $f_a$ and the inflationary Hubble scale $H$. We choose $\varepsilon = 0.3$, $m_s = H$, $m_\sigma^2 = 0.03 H^2$, and $\theta_i=1$. In the region above the dashed line $|\tilde{f}_{\rm NL}^{S,SS}|>100$ (cf.~\eqref{eq:fSSS}) and is ruled out by the current isocurvature NG searches. We also show the cooling bound on the QCD axion~\cite{Buschmann:2021juv} and bound from isocurvature power spectrum (Iso PS)~\cite{Planck:2018jri}. In the region labeled $\kappa > 1$, the restrictions~\eqref{eq:eps2},~\eqref{eq:speed_chi}, and~\eqref{eq:kin_mix} are not obeyed. We show the projected sensitivity of a search for the templates proposed in this work for two benchmark choices $\tilde{f}_{\rm NL, equi}^{\zeta\zeta S}=1, 10$ which would be able to cover the regions above the teal lines.}
 \label{fig_param}
\end{figure} 
While the above discussion has been general, we now interpret the constraints in terms of the QCD axion, which plays the role of $\chi$ in the model discussed above.
The cosmological abundance when the Peccei-Quinn symmetry is broken before inflation is given by (ignoring anharmonic effects)~\cite{Workman:2022ynf},
\es{}{
\Omega_a h^2 \approx 0.12 \left({F_a \over 9 \times 10^{11}~{\rm GeV}}\right)^{1.165} \theta_i^2.
}
The isocurvature fraction is given by,
\es{eq:iso}{
{\beta_{\rm iso}\over 1-\beta_{\rm iso}} = {P_S \over P_\zeta}= \left(\Omega_a \over \Omega_{\rm DM}\right)^2{H^2 \over \pi^2 F_a^2 \theta_i^2 {\cal P}_{\zeta}}.
}
Using Eq.~\eqref{eq:iso} to trade initial misalignment $\theta_i$ with other parameters, and saturating the current constraint $\beta_{\rm iso} < 0.038$~\cite{Planck:2018jri}, we can derive
\es{}{
H < 2.6 \times 10^{7}~{\rm GeV} \left( {0.12 \over \Omega_a h^2}\right)^{1/2} \left({F_a \over 9 \times 10^{11}~{\rm GeV}}\right)^{0.42}.
}
Requiring $|\tilde{f}^{\zeta\zeta S}_{\rm NL,equi}|>10$ implies,
\es{}{
H^3  > 10 \pi F_a \theta_i {\cal P}_{\zeta}^{1/2} \left(\Omega_{\rm DM} \over \Omega_a \right) \left(m_\sigma^2 \over 21\kappa \varepsilon \right).
}
The parameter $\kappa$ is related to the other parameters as, $\kappa = \varepsilon^2 (m_\sigma^2 \sigma_0)^2/(9H^2 m_s^2 F_a^2)$.
There is also a model-independent contribution to isocurvature NG.
This arises because the isocurvature fluctuation contains a term quadratic in the axion fluctuation,
\es{}{
S = {\Omega_a \over \Omega_{\rm DM}} \left(2 {\left(\delta a \over a\right)} + \left(\delta a \over a\right)^2\right).
}
This implies an isocurvature NG~\cite{Kawasaki:2008sn} that is independent of the kinetic mixing parameter $\varepsilon$,
\es{eq:fSSS}{
\tilde{f}_{\rm NL}^{S,SS} = {1 \over 2}{\Omega_{\rm DM} \over \Omega_a}{P_S^2 \over P_\zeta^2}.
}
Using~\eqref{eq:iso} we can project the bound on $\tilde{f}_{\rm NL}^{S,SS}$ in terms of $H$ and $f_a$.
The parameter space ruled out by the current {\it Planck} constraint on $\tilde{f}_{\rm NL}^{S,SS}$ is shown via the dashed line in Fig.~\ref{fig_param}.
We summarize the other constraints in Fig.~\ref{fig_param}.
The regions above the teal solid and dot-dashed lines would be covered by a future search of $\langle \zeta \zeta S\rangle$ NG with the shapes considered in this work.
Such a search would improve upon the $\langle S S S\rangle$ NG constraints by an order of magnitude in parts of the parameter space, and go beyond the standard isocurvature power spectrum searches for the QCD axion.

\section{Generalities}
\label{sec:general}
In the above, we have given an example in the context of the curvaton scenario in which observable isocurvature NG can arise.
It is natural to ask how general our result is, and clarify our non-minimal choice involving curvaton dynamics.
To that end, we explore two possibilities: first, a model-agnostic EFT description based on~\cite{Cheung:2007st}; second, a slow-roll inflationary scenario without the curvaton dynamics.
As we show below, we find that while the EFT description can allow a stronger interaction between the adiabatic and isocurvature fluctuation with an observable NG signal, its UV completion in a slow-roll inflaton model results in a significantly suppressed signal. It remains to be seen whether an observable signal can arise in a scenario where the scalar perturbations emerge from the inflaton.
\subsection{Goldstone EFT}
\label{sec:goldstone_eft}
In the Goldstone EFT of inflation~\cite{Cheung:2007st} the inflaton is modeled as a Goldstone boson arising from spontaneous breaking of time-translation symmetry.
We can then write an EFT for the fluctuation of the adiabatic mode, along with any other spectator field such as $\chi$, considered above.
Such an EFT including various operators up to quartic terms has been described in~\cite{Freytsis:2022aho}, building up on~\cite{Senatore:2010wk}.
Here we just select a particular cubic term to illustrate that isocurvature NG can be large and observable while being in theoretical control.
In particular, we focus on a term
\es{eq:EFT_op}{
-{1 \over \Lambda_{\rm EFT}^2} (\partial \pi_c)^2 \dot{\chi},
}
allowed by the symmetries of the theory.
The field $\pi_c$ is the canonically normalized inflaton fluctuation field and the EFT cutoff scale is denoted by $\Lambda_{\rm EFT}$, which can be below the symmetry breaking scale $f_\pi = (2\mpl^2|\dot{H}|c_s)^{1/4} \approx 60H$.
However, as long as $\Lambda_{\rm EFT} \gg H$, we can have a controlled theory of fluctuations between $H$ and $\Lambda_{\rm EFT}$.

The operator in~\eqref{eq:EFT_op} has the same structure as the one in~\eqref{eq:lag_cubic} with the replacements, $\delta\sigma \rightarrow \pi_c$, $2/(3\sigma_0) \rightarrow -H/f_\pi^2$ and $2\varepsilon^3\dot{\sigma}_0/(m_s^2 v_s^2) \rightarrow 1/\Lambda_{\rm EFT}^2$. 
Thus the associated strength of the NG is given by,
\es{}{
\tilde{f}_{\rm NL,equi}^{\zeta\zeta S} = {14 \over 3}\left(f_\pi\over H\right)^4\left({2 f_{\rm DM} H \over \chi_0}\right)\left(H^2 \over 2\Lambda_{\rm EFT}^2\right).
}
For example choosing $\Lambda_{\rm EFT} = 5 H$ and saturating the isocurvature power spectrum constraint $\beta_{\rm iso} < 0.038$, gives $\tilde{f}_{\rm NL,equi}^{\zeta\zeta S} \approx 67$ from the above.
The benchmark value considered in~\eqref{eq:tfnl_equi} corresponds to $\Lambda_{\rm EFT} \approx 18H$ in the context of this Goldstone EFT and correspondingly, the value in~\eqref{eq:tfnl_equi} is an order of magnitude smaller.

\subsection{Slow-roll EFT without the Curvaton Field}
We now consider similar NG but within the slow-roll EFT where we assume a slowly rolling inflaton field $\phi$ provides both the homogeneous expansion and the adiabatic density fluctuations.
In particular, we assume there is no curvaton field in this scenario, while there is still a spectator field $\chi$ that sources DM isocurvature.
The leading terms governing the interactions between $\phi$ and $\chi$ are given by a set of terms similar to~\eqref{eq:dim8_full},
\es{}{
{c_1 \over \Lambda^4} (\partial\phi)^4 + {c_2 \over \Lambda^4} (\partial\phi)^2 (\partial\phi\partial \chi) + {c_3 \over \Lambda^4} (\partial\phi)^2 (\partial \chi)^2 + {c_4 \over \Lambda^4} (\partial\phi \partial \chi)^2+
{c_5 \over \Lambda^4} (\partial\phi\partial \chi) (\partial \chi)^2 + {c_6 \over \Lambda^4} (\partial \chi)^4.
}
Requiring a controlled derivative expansion then forces 
\es{}{
\Lambda > \sqrt{\dot{\phi}_0},
}
from the term with the coefficient $c_1 \sim 1$.
However, adopting a simple power counting scheme where all $c_i \sim 1$, implies a small value for isocurvature NG.
To see this we can focus on the term with the coefficient $c_2$ which has the same structure as one term in~\eqref{eq:dim8_full} with the replacements $2\varepsilon^3/(m_s^2 v_s^2) \rightarrow c_2/\Lambda^4$, and $\sigma\rightarrow \phi$.
This implies,
\es{eq:sr_fnl}{
\tilde{f}_{\rm NL, equi}^{\zeta\zeta S} = {14 \over 3}{\dot{\phi}_0^2 \over H^4} \left({2 f_{\rm DM} H \over \chi_0}\right){c_2 H^2 \dot{\phi}_0 \over 2\Lambda^4}.
}
Choosing $\Lambda = 2\sqrt{\dot{\phi}_0}$ to respect the above derivative expansion bound, and saturating the isocurvature power spectrum constraint $\beta_{\rm iso} < 0.038$, gives $\tilde{f}_{\rm NL,equi}^{\zeta\zeta S} \approx 0.03$ from the above. 
This is significantly smaller than the values considered previously.
This suppression originates from requiring (1) the same field that drives the inflationary expansion also generates the adiabatic fluctuations; and (2) the same EFT cutoff for the adiabatic source and the isocurvature source.
The examples in Sec.~\ref{sec:goldstone_eft} and Sec.~\ref{sec:model} evade these restrictions by being agnostic about the origin of the homogeneous inflationary expansion and having another adiabatic source field (curvaton), respectively.

\subsection{Effects of Shift Symmetry Breaking}
One may wonder whether the suppression in the previous subsection originates from requiring shift symmetric couplings between the inflaton and $\chi$.
To explore this possibility by taking a `bottom-up' approach, we consider a non-shift symmetric (NS) operator
\es{}{
{\phi \over \Lambda_{\rm NS}}(\partial\phi\partial\chi). 
}
To have a controlled description of this interaction, the scale $\Lambda_{\rm NS}$ has to be at least bigger than the field space traversed by the inflaton in one Hubble time, i.e., $\Lambda_{\rm NS}>\dot{\phi}_0 /H$.
The induced NG is given by,
\es{}{
\tilde{f}_{\rm NL,equi}^{\zeta\zeta S} \sim {\dot{\phi}_0^2 \over H^4}\left(f_{\rm DM} H \over \chi_0\right) \left(H \over \Lambda_{\rm NS}\right).
}
Saturating $\Lambda_{\rm NS} \sim \dot{\phi}_0 /H$, this gives the same level of suppression as~\eqref{eq:sr_fnl}.
Hence even allowing for non-shift symmetric couplings, the isocurvature NG is still suppressed.

\section{Conclusion}\label{sec:con}

In this work, we highlighted the importance of NG in the bispectrum involving two adiabatic and one isocurvature perturbations. We argue that it can be naturally sizable. Due to the good sensitivities to this NG observable in CMB and LSS, it is a promising place to look for new dynamics in the early universe.  As an example, we present a simple model in which the adiabatic perturbation is generated by a curvaton. There is also an additional light scalar field responsible for generating the isocurvature perturbation, which can also be part of dark matter. Both the curvaton and the new light field have an approximate shift symmetry. We can consistently add a kinetic mixing between them which preserves all the shift symmetries. We demonstrate such a mixing can naturally arise in UV models. In the context of this model, we show that the bispectrum $\langle\zeta(\vec{k}_1)\zeta(\vec{k}_2) S(\vec{k}_3)\rangle'$ is the best observable to probe the presence of the light field and its coupling to the curvaton. This model also predicts sizable NG in bispectra involving more than one isocurvature perturbation. However, those are less useful due to limited sensitivities in CMB and LSS observations. We interpret our result in the scenario that the additional light scalar is the QCD axion, and demonstrate that the bispectrum with one isocurvature perturbation is indeed the most promising probe.

The accurate prediction of the shape of the bispectrum has an important impact on its search. The bispectrum involving one or more isocurvature perturbations has been searched for assuming the so-called local shape. In this work, we emphasize that it can naturally be of a very different shape, and demonstrate this fact by our example. We quantify the difference between the predicted shape and the local shape and find that it could be significant. Hence, using the correct shape can lead to improvement in sensitivity by at least a factor of two, and in some cases almost an order of magnitude.

There will be a large influx of data from CMB and LSS observations in the coming decades, see e.g.~\cite{Chou:2022luk} for a summary. Developing effective and comprehensive strategies will be crucial for extracting the maximal amount of information. Given the multitude of possibilities, it is essential to explore more scenarios and identify potential new signatures. Our work is a step in this direction, and much more needs to be done. For example, as discussed in Section~\ref{sec:general}, it is quite generic to expect a sizable NG signal of the kind discussed in this paper from the point of view of the effective field theory of inflation. At the same time, it remains to be seen whether this case can be UV completed in a satisfying way. Regarding complete models, we have only considered a special class, mostly motivated by enforcing shift symmetries. There are good motivations to think this is an important ingredient. At the same time, it would be interesting to explore more model building directions, which may lead to new observables and new shapes of NG. We leave all of these promising directions for future work.

\section*{Acknowledgment}
We thank J. Colin Hill and Ben Safdi for the helpful discussions.
SK is supported partially by the National Science Foundation (NSF) grant PHY-2210498 and the Simons Foundation.
LTW is supported by the Department of Energy grant DE-SC0013642. 
MG is supported by the Israel Science Foundation under Grant No. 1302/19 and 1424/23. MG is also supported by the US-Israeli BSF grant 2018236 and the NSF-BSF grant 2021779.
LTW, MG, and SK are grateful to the Aspen Center for Physics, supported by NSF grant PHY-2210452, for hospitality while this work was in progress.
We are also grateful to the Maryland Center for Fundamental Physics, University of Maryland, for hosting the Cosmological Probes of New Physics workshop where this project was initiated.

\appendix
\bibliographystyle{utphys}
\bibliography{references}
\end{document}